\documentclass[%
reprint,superscriptaddress,amsmath,amssymb, floatfix,aps,prc,twocolumn]{revtex4-2}

\usepackage{graphicx}
\usepackage{caption}
\usepackage{subcaption}
\usepackage{bm}
\usepackage{amsmath}   

\usepackage{orcidlink}
\usepackage{hyperref}
\hypersetup{breaklinks=true, colorlinks=true, citecolor=blue}
\usepackage{color}
\usepackage[normalem]{ulem}  

\newcommand{\pt}[1]{\textsf{\color[rgb]{0,0.4,0}{ #1}}}

\newcommand{\MeV}{\text{MeV}}
\newcommand{\G}{\text{G}}
\newcommand{\kHz}{\text{kHz}}
\newcommand{\ms}{\text{ms}}

\newcommand{\NWA}{\text{NWA}}
\newcommand{\ndUrca}{\ensuremath{n_\text{dU}}}
\newcommand{\dUrca}{\text{dU}}
\newcommand{\muI}{\Delta\mu}

\begin{document}


\title{Thermal and Magnetic Effects on Bulk Viscosity in Binary Neutron Star Mergers}

\author{Pranjal Tambe, \orcidlink{0000-0003-2293-6953}}
 \email{pranjal.tambe@iucaa.in}
 \affiliation{%
 Inter University Centre for Astronomy and Astrophysics, Ganeshkind, Pune 411007, India
}%

\author{Debarati Chatterjee, \orcidlink{0000-0002-0995-2329}}%
 \email{debarati@iucaa.in}
\affiliation{%
 Inter University Centre for Astronomy and Astrophysics, Ganeshkind, Pune 411007, India
}%

\author{Mark Alford, \orcidlink{0000-0001-9675-7005}}
\affiliation{%
 Physics Department, Washington University, St.~Louis, MO 63130, USA
}%

\author{Alexander Haber, \orcidlink{0000-0002-5511-9565}}
 \email{ahaber@physics.wustl.edu}
\affiliation{%
 School of Mathematical Sciences and STAG research centre, University of Southampton, Hampshire, UK
}%


\begin{abstract}
Astrophysical scenarios such as binary neutron star mergers, proto-neutron stars and core-collapse supernovae involve finite temperatures and strong magnetic fields. Previous studies on the effect of magnetic fields on flavor-equilibration processes relied on the Fermi surface approximation, which is not a reliable 
approximation in the neutrino-transparent regime of matter in supernovae or neutron star mergers. In a recent study, we went beyond the Fermi surface approximation, performing the full phase space integral to obtain direct Urca rates in a background magnetic field. In this work, we extend these calculations to incorporate the collisional broadening (``modified Urca'') contribution. We use the recently developed Nucleon Width Approximation, which naturally includes the magnetic field dependence of all contributions. We demonstrate the impact of magnetic
fields on the flavor-equilibrium condition for two finite-temperature equations of state with different direct Urca thresholds. We also study the impact of magnetic fields on the bulk viscous dissipation of density oscillations relevant in postmerger scenarios. 
\end{abstract}

\maketitle
\section{Introduction}
\label{sec:intro}
Neutron Stars (NS) are compact astrophysical objects that contain the densest form of matter in the universe. This makes them ideal laboratories to probe the properties of matter at densities beyond nuclear saturation density $n_0$ and where the matter is highly isospin asymmetric~\cite{Glendenning,Schaffner-Bielich_2020}.
Simulations of explosive events such as neutron star mergers and supernovae require predictions of the dynamical properties of
nuclear matter. The properties we explore in this work are related to the equilibration of flavor, specifically the rate of ``Urca'' processes, i.e., those that convert neutrons into protons and vice versa. 
In the violent dynamics of binary neutron star mergers and supernovae we expect high-amplitude density oscillations in the kHz frequency range that drive the nuclear matter
out of flavor equilibrium
\cite{Janka:2006fh,Lentz:2015nxa,Baiotti:2016qnr, Alford:2017rxf,Hanauske:2019qgs, Burrows:2020qrp, Most:2022wgo}.
Previous work has shown that the re-equilibration
of flavor (i.e., beta equilibration of the proton fraction) via Urca processes can produce
strong bulk viscous damping of these oscillations \cite{Alford:2017rxf, Alford:2019qtm, Alford:2020lla, Most:2021zvc, Alford:2021lpp, Alford:2022ufz, Most:2022yhe, Alford:2023uih}.
Urca rates are also important for calculations of the cooling of hot proto-neutron stars and merger remnants via neutrino emission.
The main thrust of this paper is to explore 
the effect of magnetic fields on Urca rates.

Neutron stars have been observed to have strong magnetic fields ($B \approx 10^9{-}10^{12}\,\G$). Magnetars have been observed to have ultra-strong surface magnetic fields as large as $B\approx10^{15}\,\G$ and could potentially have even higher interior magnetic fields \cite{Mereghetti:2015asa, Kaspi:2017fwg, Konar:2017kty, Esposito:2018gvp}. Moreover, merger simulations indicate that magnetic fields are amplified during the merger and the remnant can be strongly magnetized \cite{Harding:2006qn, Lorimer:2008se, Kiuchi:2015sga, Ciolfi:2017uak, Ciolfi:2020cpf, Aguilera-Miret:2020dhz, Aguilera-Miret:2025nts, Neuweiler:2025klw, Reboul-Salze:2024jst}. Recent calculations, e.g.~see \cite{Aguilera-Miret:2020dhz, Aguilera-Miret:2025nts, Neuweiler:2025klw}, show that the rapid amplification of the magnetic field to volume averaged values of $10^{16}$ G is a robust outcome of BNS merger simulations, and support the presence of locally higher magnetic fields exceeding $10^{17}$ G. It has been shown in \cite{Reboul-Salze:2024jst} that the Taylor-Spruit dynamo mechanism can amplify magnetic fields to $10^{17}$ G in the post-merger remnant, which is a hyper-massive neutron star. Increasing resolution may further increase the field.
It is therefore important to consistently include magnetic field effects when calculating the properties of nuclear matter for inclusion in simulations of mergers or supernovae.

In our calculations, we assume that nuclear matter is transparent to neutrinos, i.e., the neutrino mean free path is not much smaller than the size of the neutron star. This is a reasonable approximation at temperatures $T\lesssim 10\,\MeV$ \cite{ Sawyer:1975js, Sawyer:1978qe, Haensel:1987zz, Roberts:2016mwj,Alford:2018lhf}. Such temperatures are easily attained in 
supernovae \cite{Pons:1998mm, Janka:2006fh, Lentz:2015nxa, Burrows:2020qrp} and binary neutron star mergers \cite{Hanauske:2019qgs, Most:2022wgo}.
In this regime the relevant Urca processes  are neutron decay (nd) and electron capture (ec):
\begin{align}
 n&\to p+e^-+\bar\nu_e \quad\text{(nd)} \ , \label{eq:nd-def}\\
 p+e^- &\to n+\nu_e \qquad\quad\text{(ec)} \ . \label{eq:ec-def}
\end{align}

The standard approach to calculating Urca rates has been to distinguish two contributions, the ``direct Urca'' (Eqs.~\ref{eq:nd-def} and~\ref{eq:ec-def}), which are dominant above a threshold density but (at $T\ll 1\,\MeV$) strongly suppressed below that density by the momentum conservation condition, and the ``modified Urca'', which always satisfies the momentum conservation condition in presence of a bystander nucleon and contributes both below and above the direct Urca threshold density.
In the low temperature limit, where one can simplify the phase space integral using the Fermi surface approximation, it is straightforward to calculate the effect of strong magnetic fields on these rates (see, e.g., \cite{Lai1991, Leinson:1998yr, Baiko:1998jq, Yuan:1998, Anand:2000ud, Sinha:2008wb, Zhang:2010zzk, Maruyama:2021ghf}.
However, the Fermi surface approximation becomes invalid at temperatures $T\gtrsim 1\,\MeV$ \cite{Alford:2018lhf}. One must then perform the full phase space integral, including magnetic field effects.
In our recent work \cite{Tambe:2024usx}, we performed the calculation for this direct Urca contribution to the rates, which dominates at densities above the direct Urca threshold. In this paper, we take the next step, including magnetic field effects on the full Urca rate, at densities below and above the threshold.

To do this, we use a recent reformulation of the Urca rate calculation, the nucleon width approximation (NWA) \cite{Alford:2024xfb}. In the NWA framework, the modified Urca contribution, which arises from collisional broadening of the in-medium nucleon states, is naturally included on the same footing as the direct Urca contribution, and it is straightforward to incorporate magnetic fields into the whole framework.

This paper is structured as follows. 
In Sec.~\ref{sec:formalism} we present the NWA formalism for the calculation of rates of Urca processes in the presence of magnetic fields. In Sec.~\ref{sec:results} we show the results for the Urca process rates and the bulk viscosity for the two equations of state (EoS) used in this paper. In Sec.~\ref{sec:discussions} we discuss the implications of our findings. We use natural units where $\hbar=c=k_B=1$ throughout this paper.

\begin{figure}
    \centering
    \includegraphics[width=0.45\textwidth]{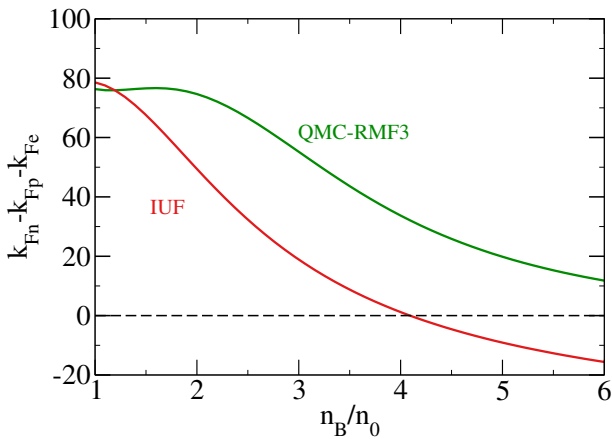}
    \caption{Momentum deficit $k_\text{Fn}-k_\text{Fp}-k_\text{Fe}$ for the IUF and QMC-RMF3 EoS at $T=0$ MeV. Direct Urca is kinematically suppressed for positive values of the deficit. }
    \label{fig:momshort}
 \end{figure}

\section{Formalism}
\label{sec:formalism}

Our aim is to calculate the Urca process rates in the presence of a magnetic field above and below the direct Urca threshold density $n_B=\ndUrca$, at temperatures $T\gtrsim 1\,\MeV$ where the full phase space integral must be evaluated, but $T\lesssim 10\,\MeV$ where our assumption of neutrino transparency is reasonable. We consider magnetic fields as low as $5 \times 10^{16}$ G to as high as $5 \times 10^{17}$ G. For the lowest field considered, the magnetic field does not modify the EoS and one can consider a zero field EoS. For the highest field considered, there will be Landau quantization effects on the EoS, but the inclusion of these would not change the results significantly. This can be seen from 
Appendix~\ref{app:landauquant}, where we compare the calculations of bulk viscosity for QMC-RMF3 EoS  at $T=1\,\MeV$ with and without the inclusion of Landau quantization in the EoS. We see that the amplitude of the de Haas–van Alphen oscillations is enhanced if we use an EoS with finite temperature and finite magnetic field effects, but the oscillation-averaged value of bulk viscosity remains unaffected. As the magnetic field effects are washed out at higher temperatures, the difference will be negligible for higher temperatures.

As noted in Sec.~\ref{sec:intro}, we use the NWA formalism, which naturally includes both direct Urca and modified Urca contributions in a single expression where it is straightforward to incorporate magnetic fields.
The NWA expression for the Urca rate is:
\begin{equation}
    \Gamma^{\NWA} = \int_{-\infty}^{\infty} dm_n\, dm_p\, \Gamma^{\dUrca}(m_n,m_p)\,R_n(m_n)\,R_p(m_p)\, , 
\label{eq:Rates_NWA}
\end{equation}
where $\Gamma^{\dUrca}$ is the direct Urca rate and the mass-spectral function for nucleon $N\in{n,p}$ is
\begin{equation}
    R_N(m) = \frac{1}{\pi} \frac{W_N/2}{(m - M_N^*)^2 + W_N^2/4}\, .
\label{eq:mass_spectrum}
\end{equation}

Here, $M_N^*$ is the in-medium effective nucleon mass and $W_N$ is the nucleon width, which captures the collisional broadening of the nucleon state arising from strong interactions with the surrounding nuclear medium. In the simplest implementation of  NWA, the effective mass and the width are taken to be independent of the nucleon's momentum and energy, depending only on the density and temperature of the surrounding nuclear medium.

In this paper, we use our previous calculation of direct Urca rates in the presence of a magnetic field \cite{Tambe:2024usx}, combined with Eq.~\eqref{eq:Rates_NWA} and an estimate of the nucleon width 
\begin{equation}
    W_N = \dfrac{T^2}{T_W},\quad T_W=5\,\MeV \ ,
    \label{eq:width}
\end{equation}
obtained from a Brueckner theory calculation for pure
neutron matter using the Paris NN potential in Ref.~\cite{Sedrakian:2000kc}, to obtain the Urca process rates. The neutron decay direct Urca rate in the presence of a magnetic field is \cite{Tambe:2024usx}:
\begin{align}
    \Gamma^{\dUrca}_\text{nd}=&\frac{G^2(1+3g_A^2)eB}{16\pi^5} \int \,dk_n\, dk_{pz}\, dk_{ez}\, E_{\nu}^2\, k_n\notag\Theta(E_{\nu})\\& \Theta(k_n-\lvert k_{pz}+k_{ez}\rvert)\,\sum_{l,l'} (F^2_{l',l}(u)+F'^2_{l',l-1}(u)) \notag\\& f_n (1-f_p)(1-f_e) \,\label{eq:Rates_mag},    
\end{align}
where $E_{\nu}=E_n-E_p-E_e$ is the neutrino energy fixed by energy conservation and $f_i = \left[1 + \exp((E_i-\mu_i)/T)\right]^{-1}$ are the Fermi-Dirac distribution functions for the constituent particles. For a definition of all relevant quantities, see Ref.~\cite{Tambe:2024usx}. The functions $F_{l',l}$ are the normalized Laguerre functions,
\begin{equation}
    F_{l',l}(u) = \sqrt{\frac{l'!}{l!}} u^{(l-l')/2} e^{{-u}/{2}} L_{l'}^{l-l'}(u)\, = (-1)^{l'-l} F_{l,l'}(u) ,
    \label{eq:lag_f}
\end{equation}
where $L_{l'}^{l-l'}(u)$ are the Laguerre polynomials and $l, l'$ are indices for the electron and proton Landau level numbers~\cite{Tambe:2024usx}.
 The rate for the electron capture process has a similar expression with phase space factors $f_n (1-f_p)(1-f_e)$ replaced by $(1-f_n) f_p\,f_e$ and $E_{\nu} = E_p+ E_e - E_n$.

To describe the nuclear matter, we use two different relativistic mean field models that satisfy current astrophysical and nuclear constraints: IUF \cite{Fattoyev:2010mx} and QMC-RMF3 \cite{Alford:2022bpp, Alford:2023rgp}. 
The IUF equation of state (EoS) has a direct Urca threshold at
$n_\text{dU} = 4.1 n_0$, while the QMC-RMF3 EoS has no direct Urca threshold in the range of densities considered in our study. We show this explicitly in
Fig.~\ref{fig:momshort}, where we plot the
direct Urca  momentum deficit $k_{F_n}-k_{F_p}-k_{F_e}$ for both EoSs. When the deficit is positive, direct Urca is kinematically forbidden for degrees of freedom near their Fermi surface, so the direct Urca process is exponentially suppressed at low temperature (see Appendix~\ref{app:estimating}). We restrict our calculations to $npe$ matter; extending the calculations to include muons would be straightforward but would not make a significant difference
(see, e.g., \cite{Alford:2022ufz}). 

One of the physical quantities we determine is the flavor equilibrium condition. Flavor equilibrium is established when the rates of proton creation (neutron decay, Eq.~\eqref{eq:nd-def}) and proton destruction (electron capture, Eq.~\eqref{eq:ec-def}) are equal, $\Gamma_\text{nd}=\Gamma_\text{ec}$. 
Since the neutrinos are not thermally equilibrated, and reactions \eqref{eq:nd-def} and \eqref{eq:ec-def} are not time-inverses of each other, equilibrium between them is not in general determined by a simple balancing of chemical potentials. 
At low temperatures $T\ll 1\,\MeV$, the neutrinos can be neglected since they carry away a very small amount of energy compared to the Fermi energies of the participating particles, in which case the processes {\em are} time-inverses of each other and we can use the principle of detailed balance to obtain the cold flavor equilibrium condition, $\mu_n=\mu_p + \mu_e$.
In the temperature range $1\,\MeV \lesssim T \lesssim 10\,\MeV$ that is of interest to us, however, this is no longer valid since
the neutrinos, while still free-streaming, now carry away a significant amount of energy. To achieve the equilibrium condition $\Gamma_\text{nd}=\Gamma_\text{ec}$ we now require a mismatch between $\mu_n$ and $\mu_p+\mu_e$, which
we express as an isospin chemical potential $\muI$,
\begin{equation}
    \mu_n=\mu_p+\mu_e+\muI \ .
    \label{eq:muI-def}
\end{equation}
For a detailed discussion of this phenomenon at vanishing magnetic field, see Refs.~\cite{Alford:2018lhf,Alford:2021ogv}. In this work,
we calculate $\muI$ including its dependence on the magnetic field.

We also evaluate the bulk viscosity arising from re-equilibration of the proton fraction in nuclear matter subjected to a small-amplitude density oscillation of angular frequency $\omega$. This is given by \cite{Alford:2023gxq}:
\begin{equation}
    \zeta =  \frac{\partial p}{\partial x_I} \Bigg\vert_{n_B} \frac{\gamma_B}{\gamma_I^2 + \omega^2} \, ,
    \label{eq:bvisc}
\end{equation}
where
\begin{align}
    \gamma_I &\equiv - \frac{1}{n_B} \frac{\partial\Gamma_I}{\partial x_I}\Bigg\vert_{n_B} ,\label{eq:relaxrate}\\
    \gamma_B &\equiv \frac{\partial\Gamma_I}{\partial n_B}\Bigg\vert_{x_I} .
\end{align}
Here $\Gamma_I$ is the net rate per unit volume at which isospin $I_3$ increases and $x_I$ is the isospin fraction related to the proton fraction $x_p$ by $x_I=x_p- \frac{1}{2}$. The isospin relaxation rate is $\gamma_I$, and the
isospin relaxation time is $\tau_I = 1/\gamma_I$. At low temperatures $T\ll 1\,\MeV$, where equilibrium is characterized by the isospin chemical potential $\mu_I$ being zero, $\gamma_B$ is not an independent quantity, it is proportional
to $\gamma_I$ (see \cite{Alford:2023gxq}, Appendix A). 
Our calculations explore temperatures up to 5\,\MeV, where, because neutrinos carry significant energy but are not in thermal equilibrium, $\mu_I$ is not zero in flavor equilibrium \cite{Alford:2018lhf}, and then
$\gamma_B$ must be calculated separately. A quantity related to bulk viscosity that is relevant in binary neutron star mergers is the damping time of density oscillations, given by
\begin{equation}
    \tau_\text{damp}=\frac{\kappa^{-1}}{\omega^2\zeta(\omega)},
    \label{eq:tdamp}
\end{equation}
where $\kappa^{-1}$ is the incompressibility calculated from the EoS, given by
\begin{equation}
    \kappa^{-1}=n_B\frac{\partial p}{\partial n_B}\Bigg\vert_{x_I ,\,T}.
\end{equation}

 Our goal here is to show that flavor equilibration via weak interactions (a) has a noticeable effect, and (b) needs to have its rates calculated properly, including the thermal and magnetic field effects that we discuss here.  To illustrate that point, we perform calculations in the linear response regime, which allows us to express the physical consequences in simple terms (bulk viscosity, damping time, etc.). This is intended as motivation for including these weak interaction rates in merger simulations.  Numerical simulations suggest that during the first few milliseconds following merger, density oscillations can reach large amplitudes, in which case the suprathermal bulk viscosity would be relevant \cite{Madsen:1992sx,Reisenegger:2003pd}. As demonstrated in Refs.~\cite{Most:2022yhe,Camelio:2022ljs,Camelio:2022fds,Gavassino:2023xkt}, a direct implementation of the Urca reaction rates in merger simulations naturally captures the full physical behavior, encompassing both the subthermal and suprathermal bulk viscosity effects. 

\begin{figure}
 \centering
 \includegraphics[width=0.45\textwidth]{IUF_T1_ND.eps}
 \caption{Neutron Decay rates for IUF matter at $T=1\,\MeV$, comparing direct Urca at magnetic field  $B=0$ and NWA result for various values of $B$. The black dashed line marks the direct Urca threshold for the IUF EoS at $T=0$ and $B=0$.
  }
    \label{fig:nd_IUF_T1_finiteB}
\end{figure}

\begin{figure*}
    \begin{subfigure}{0.33\textwidth}
        \centering
    \includegraphics[width=\textwidth]{IUF_T1_ND_finiteB.eps} 
    \label{fig:IUF_T1_ND_finiteB}
    \end{subfigure}\hfill
    \begin{subfigure}{0.33\textwidth}
        \centering
    \includegraphics[width=\textwidth]{IUF_T3_ND_finiteB.eps} 
    \label{fig:IUF_T3_ND_finiteB}
    \end{subfigure}\hfill
    \begin{subfigure}{0.33\textwidth}
         \centering
    \includegraphics[width=\textwidth]{IUF_T5_ND_finiteB.eps}
    \label{fig:IUF_T5_ND_finiteB}
    \end{subfigure}
\caption{
Neutron decay rates at three temperatures, $T=1,3,5\,\MeV$, and magnetic fields $B=10^{17}, 3\times10^{17},$ and $5\times10^{17}\,\G$ using the IUF EoS. }
\label{fig:IUF_ND_finiteB}
\end{figure*}

\begin{figure*}
    \begin{subfigure}{0.33\textwidth}
        \centering
    \includegraphics[width=\textwidth]{IUF_T1_EC_finiteB.eps} 
    \label{fig:IUF_T1_EC_finiteB}
    \end{subfigure}\hfill
    \begin{subfigure}{0.33\textwidth}
        \centering
    \includegraphics[width=\textwidth]{IUF_T3_EC_finiteB.eps} 
    \label{fig:IUF_T3_EC_finiteB}
    \end{subfigure}\hfill
    \begin{subfigure}{0.33\textwidth}
         \centering
    \includegraphics[width=\textwidth]{IUF_T5_EC_finiteB.eps}
    \label{fig:IUF_T5_EC_finiteB}
    \end{subfigure}
\caption{
Electron capture rates at three temperatures, $T=1,3,5\,\MeV$, and magnetic fields $B=10^{17}, 3\times10^{17},$ and $5\times10^{17}\,\G$ using the IUF EoS.}
\label{fig:IUF_EC_finiteB}
\end{figure*}

\begin{figure*}
    \begin{subfigure}{0.32\textwidth}
        \centering
    \includegraphics[width=\textwidth]{dmu_IUF_T1.eps} 
    \end{subfigure}\hspace{0.01\hsize}%
    \begin{subfigure}{0.32\textwidth}
        \centering
    \includegraphics[width=\textwidth]{dmu_IUF_T3.eps} 
    \end{subfigure}\hspace{0.01\hsize}%
    \begin{subfigure}{0.32\textwidth}
         \centering
    \includegraphics[width=\textwidth]{dmu_IUF_T5.eps}  
    \end{subfigure}
\caption{
Correction $\muI$ to the flavor equilibrium condition at three temperatures, $T=1,3,5\,\MeV$, using the NWA rates for IUF EoS. We see that $\muI$ decreases with increasing magnetic field. }
\label{fig:IUF_dmu}
\end{figure*}

\begin{figure*}
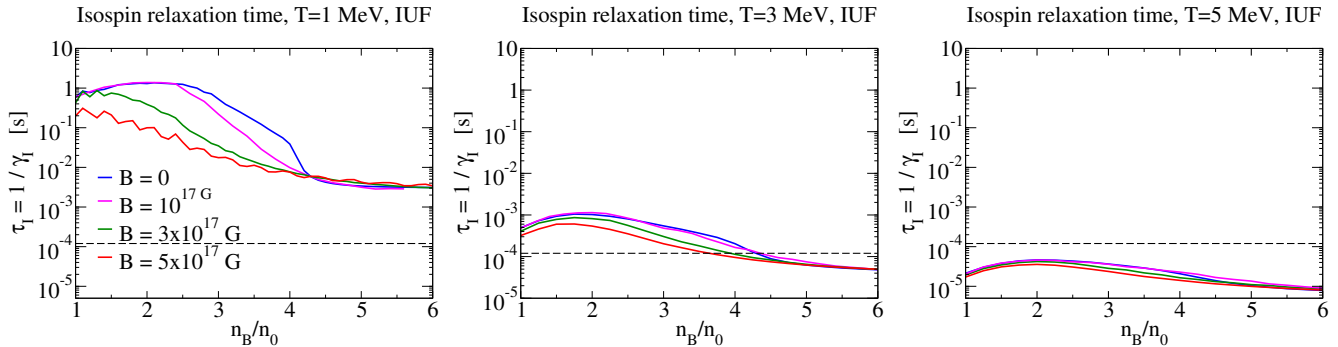

    \begin{subfigure}{0.32\textwidth}
    \includegraphics[width=\textwidth]{tflavor_T1_IUF.eps}
     \end{subfigure}\hspace{0.01\hsize}%
    \begin{subfigure}{0.32\textwidth}
    \includegraphics[width=\textwidth]{tflavor_T3_IUF.eps}
    \end{subfigure}\hspace{0.01\hsize}%
    \begin{subfigure}{0.32\textwidth}
    \includegraphics[width=\textwidth]{tflavor_T5_IUF.eps}
    \end{subfigure}
    \caption{
    Isospin relaxation time $\tau_I=1/\gamma_I$ for IUF EoS at $T=1,3,5\,\MeV$, showing how, particularly at lower temperature, magnetic field speeds up equilibration at below-threshold densities. The dashed black line corresponds to $\tau_I=0.16$ ms, which is equal to $1/\omega$ for oscillations of frequency $f=1$ kHz ($\omega=2\pi f$).}
    \label{fig:tflavor}
    
\end{figure*}

\begin{figure*}
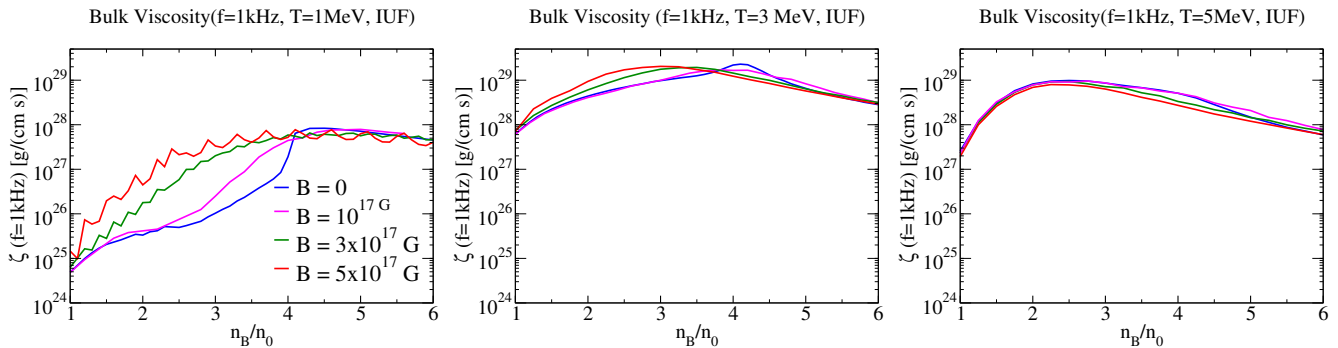

    \begin{subfigure}{0.32\textwidth}
        \centering
    \includegraphics[width=\textwidth]{bvisc_T1_IUF.eps}
    \label{fig:bvisc_T1_IUF}
    \end{subfigure}\hspace{0.01\hsize}%
    \begin{subfigure}{0.32\textwidth}
        \centering
    \includegraphics[width=\textwidth]{bvisc_T3_IUF.eps}
    \label{fig:bvisc_T3_IUF}
    \end{subfigure}\hspace{0.01\hsize}%
    \begin{subfigure}{0.32\textwidth}
        \centering
    \includegraphics[width=\textwidth]{bvisc_T5_IUF.eps}
    \label{fig:bvisc_T5_IUF}
    \end{subfigure}
    \caption{Bulk Viscosity for a $1\,\kHz$ density oscillation in matter described by the IUF EoS, at $T=1,3,5$ MeV.
    }
    \label{fig:bvisc_IUF}
\end{figure*}

\begin{figure}
        \centering        \includegraphics[width=0.45\textwidth]{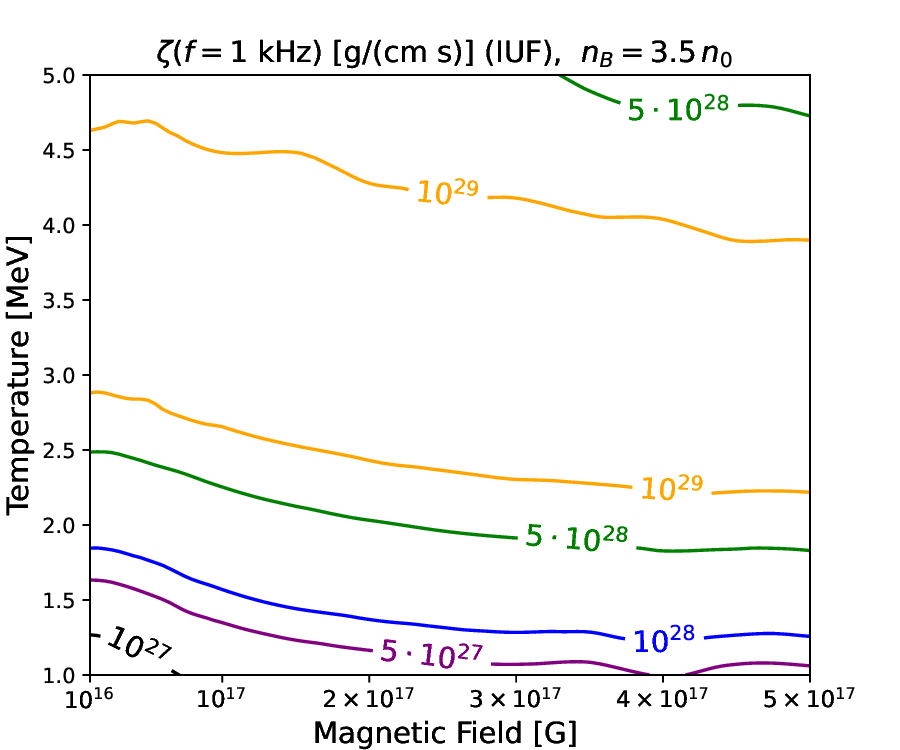}
    \caption{Contours for Bulk Viscosity  for IUF EoS as a function of temperature and magnetic field at density $n_B=3.5n_0$ (below the direct Urca threshold density).
    \label{fig:bv_contour}
    }
\end{figure}

\begin{figure}
        \centering        \includegraphics[width=0.45\textwidth]{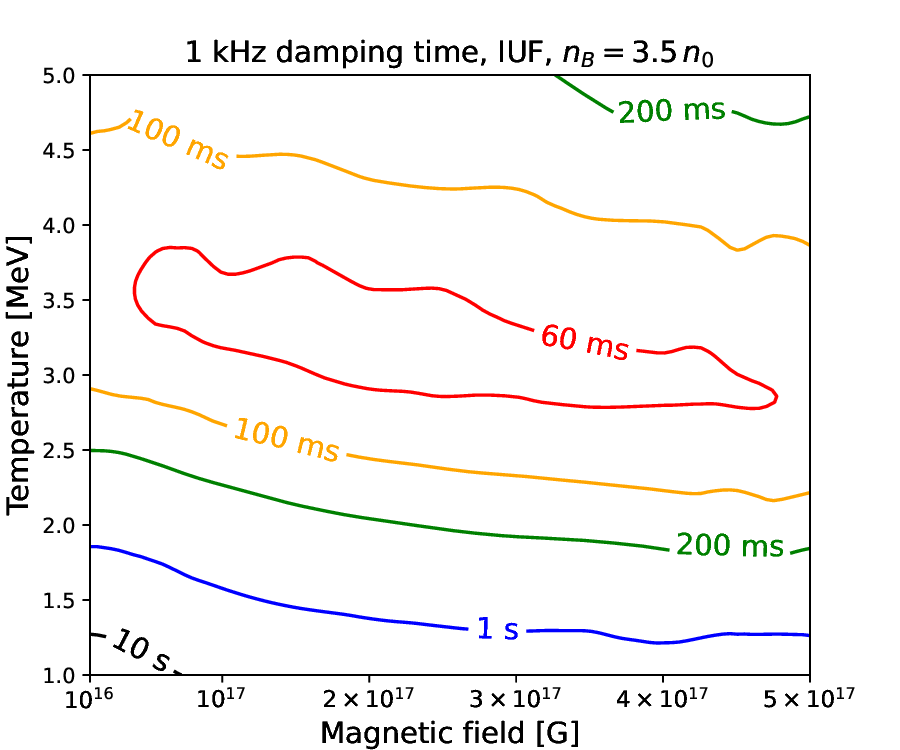}
    \caption{Contours for damping time of density oscillations of frequency $1$ kHz for IUF EoS as a function of temperature and magnetic field at density $n_B=3.5n_0$ (below the direct Urca threshold density).}
    \label{fig:tdamp_contour}
\end{figure}

\section{Results}
\label{sec:results}
As noted above, we perform calculations for
two  finite-temperature relativistic mean field EoSs, IUF (with direct Urca threshold at $n_B=4.1 n_0$) and QMC-RMF3 (no direct Urca threshold). 

\subsection{IUF}
\label{sec:results-IUF}
We compute the neutron decay and electron capture rates in the NWA formalism for IUF matter in non-vanishing magnetic fields. The neutron decay rates at $T=1\,\MeV$ are plotted in Fig.~\ref{fig:nd_IUF_T1_finiteB}.
We compare our rate calculations in the presence of a magnetic field with those in the absence of any magnetic field (the blue solid line). We highlight the difference between these calculations. For simplicity, the calculations in presence of magnetic field have been performed by neglecting the neutrino momentum in the momentum conserving $\delta$-function ($\delta(k_{n_z}-k_{p_z}-k_{e_z}-k_{\nu _z})$) \pt{\cite{Tambe:2024usx}}, while there is no such assumption in rate calculations in the absence of a magnetic field \cite{Alford:2018lhf}. This implies that the low B curves do not perfectly converge to the $B=0$ limit. However, we show the $B=0$ case to contextualize the magnitude of the impact of the magnetic field. The matrix element used in both the calculations is the same. The direct Urca rate (blue line, shown for $B=0$) includes the effects of thermal blurring of the Fermi surfaces, but this only opens up an exponentially suppressed amount of phase space (see the discussion leading to Eq.~\ref{eq:R-est-thermal}), so the rate remains heavily suppressed.

We first discuss the behavior of the rates at densities below the direct Urca threshold.
In this regime, the total rate in zero magnetic field (blue line) is dominated by the collisional broadening (``modified Urca'') contribution, where the width of the in-medium nucleons opens up some phase space for the Urca process to proceed. By utilizing the NWA framework, we account for this effect.
As the magnetic field is increased, the separation into Landau levels opens up additional phase space. We see this effect becoming important as the field rises above $5\times 10^{16}\,\G$. This is consistent with the expectation based on comparing Eqs.~\eqref{eq:R-est-NWA} and \eqref{eq:R-est-mag}.
We see that a sufficiently strong magnetic field enhances the neutron decay rate by an order of magnitude near the direct Urca threshold density.  At magnetic fields above $10^{17}$ G, we notice the appearance of de Haas van Alphen oscillations. Above the direct Urca threshold, the rates roughly converge to the zero field limit.

To analyse this further, in Figs.~\ref{fig:IUF_ND_finiteB} and \ref{fig:IUF_EC_finiteB} we plot the neutron decay and electron capture rates in the NWA formalism for varying magnetic fields at $T=1,3,5\,\MeV$. At each density we impose the condition $\mu_n=\mu_p+\mu_e$, which would correspond to beta equilibrium if the temperature were well below $1\,\MeV$.
Our previous work \cite{Tambe:2024usx} showed that when evaluating the effects of magnetic fields on direct Urca rates near or below the direct Urca threshold, the neutron decay rate is more sensitive to the magnetic field than the electron capture rate.  We can now confirm that this is also true when calculating the Urca rate consistently, which we do using the NWA formalism. 
We further observe that at higher temperatures, the effect of the magnetic field is less pronounced. This is because as the temperature rises, both thermal blurring of the Fermi surfaces and collisional broadening of the in-medium nucleons help to open up more phase space, and the magnetic field only has a noticeable effect if it makes a comparable contribution. The relative sizes of these effects are estimated in Appendix~\ref{app:estimating}.

As described in Sec.~\ref{sec:formalism} the difference between forward (proton creation, i.e.~neutron decay) and backward (proton destruction, i.e.~electron capture) rates, which is biggest near the direct Urca threshold,
leads to a finite-temperature correction $\muI$ to the flavor-equilibrium condition (Eq.~\eqref{eq:muI-def}). In Fig.~\ref{fig:IUF_dmu} we show plots of $\muI$ as a function of density  for $T=1,3,5\,\MeV$ and varying magnetic fields. We note the following features:\\
(1) $\muI$ is largest at densities just below the direct Urca threshold $n_B=4.1n_0$; \\
(2) Increasing temperature increases $\muI$; \\
(3) Increasing magnetic field generally decreases $\muI$.

To understand these features, we refer to
the general trends seen in Figs.~\ref{fig:IUF_ND_finiteB} and \ref{fig:IUF_EC_finiteB}, which show the rates
in the absence of the $\muI$ correction, at
$\mu_n=\mu_p+\mu_e$. \\
(1) Thermal blurring spreads out the direct Urca threshold over a range of densities: looking at the $B=0$ lines just below the direct Urca threshold we see that electron capture is enhanced more than neutron decay. For an explanation in terms of the differing kinematics, see Ref.~\cite{Alford:2018lhf}. This greater mismatch between the rates means that a larger value of $\muI$ is needed to bring the rates together to balance each other. \\
(2) Since thermal blurring is responsible for the rate mismatch, higher temperatures mean a larger mismatch and hence a larger value of $\muI$.\\
(3) We see that as the magnetic field increases, the forward and backward rates become more similar. Thus we expect that near the direct Urca threshold density the correction $\Delta\mu$ should decrease with increasing magnetic field.

Urca processes are the mechanism by which the proton fraction in nuclear matter relaxes to its beta equilibrium value. In Fig.~\ref{fig:tflavor}
we plot the isospin relaxation time \mbox{$\tau_I=1/\gamma_I$} in for temperatures $T=1,3,5\,\MeV$. Focusing on densities below the direct Urca threshold, we see that at $T=1\,\MeV$, where thermal contributions to the rate are small, magnetic fields in the $10^{17}\,\G$ range significantly increase the rate and thereby shorten the relaxation time. 
However, as the temperature rises to $5\,\MeV$, the relaxation time at $B=0$ shortens as thermal effects (thermal blurring of the Fermi surfaces and collisional broadening of the in-medium nucleons) overcome the kinematic barrier to direct Urca. At these higher temperatures, there is little additional effect of magnetic fields in the $10^{17}\,\G$ range.

One of the physical manifestations of isospin relaxation is bulk viscosity, which damps density oscillations. Bulk viscosity attains a
resonant maximum when the relaxation rate $\gamma_I$ matches the angular frequency $\omega$ of the density oscillation. The oscillations seen in merger simulations typically have frequency $f  \approx 1$ kHz, so maximum bulk viscosity is reached when $\tau_I\approx 0.16\,\ms$. We see already from Fig.~\ref{fig:tflavor} that this occurs at densities up to the direct Urca threshold and temperatures of about $2$ to $4\,\MeV$ and that at $T\lesssim 3\,\MeV$ magnetic fields of order $10^{17}\,\G$ can shift the density or temperature at which the resonant maximum occurs.

To see this effect more clearly, in  Fig.~\ref{fig:bvisc_IUF} we plot the bulk viscosity from Eq.~\eqref{eq:bvisc} for density oscillations with $f=1\,\kHz$ ($\omega= 2\pi{\times}1\,\kHz$) for different magnetic fields, at temperatures $T=1,3,5\,\MeV$.
At  $T=1\,\MeV$ the relaxation rate is slower than $2\pi{\times}1\,\kHz$, so the enhancement of the rate by the magnetic field leads to an increase in bulk viscosity. At  $T=5\,\MeV$ the relaxation rate is faster than $2\pi{\times}1\,\kHz$, so the slight enhancement of the rate by the magnetic field leads to a small decrease in bulk viscosity.
At $T=3\,\MeV$ the rate passes through $2\pi{\times}1\,\kHz$ as the density is varied, so we see a resonant peak, which shifts to lower density as the magnetic field is raised.

In Fig.~\ref{fig:bv_contour} we use a contour plot to show the combined dependence of the bulk viscosity on temperature and magnetic field,
for the IUF EoS at density $n_B=3.5\,n_0$ which is below the direct Urca threshold. We see that the bulk viscosity has a maximum (in the region bounded by the orange contours) that moves down to lower temperature as the magnetic field increases, thereby enhancing the rate, so that resonance with $f=1\,\kHz$ is achieved at lower temperatures. 

Finally, to give a sense of how the effects discussed above are manifested in the behavior of nuclear matter, we show in Fig.~\ref{fig:tdamp_contour}
a plot of the damping time of density oscillations with frequency $f=1\,\kHz$ as a function of temperature and magnetic field,
again at density $n_B=3.5\,n_0$ below the direct Urca threshold. The
region of fastest damping is within the orange $60\,\ms$ contour, which moves to lower temperatures as the magnetic field increases. This is what one would expect given that this is where the bulk viscosity reaches its maximum (see Fig.~\ref{fig:bv_contour}).

 \begin{figure}
    \centering
    \includegraphics[width=0.45\textwidth]{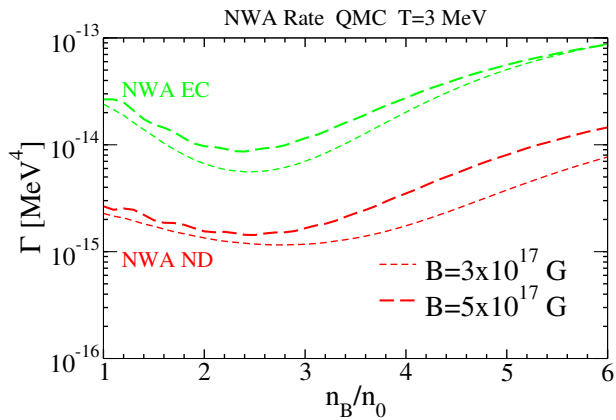}
    \caption{Neutron decay and electron capture rates in the NWA formalism for the QMC-RMF3 EoS at $T=3\,\MeV$ in cold chemical equilibrium. 
    }
    \label{fig:QMC_T3_finiteB}
\end{figure}

\begin{figure*}
    \begin{subfigure}{0.32\textwidth}
        \centering
    \includegraphics[width=\textwidth]{dmu_QMC_T1.eps} 
    \end{subfigure}\hspace{0.01\hsize}%
    \begin{subfigure}{0.32\textwidth}
        \centering
    \includegraphics[width=\textwidth]{dmu_QMC_T3.eps} 
    \end{subfigure}\hspace{0.01\hsize}%
    \begin{subfigure}{0.32\textwidth}
         \centering
    \includegraphics[width=\textwidth]{dmu_QMC_T5.eps}  
    \end{subfigure}
\caption{
Correction $\muI$ to the flavor equilibrium condition at three temperatures, $T=1,3,5\,\MeV$, using the NWA rates for QMC-RMF3 EoS. We see that $\muI$ decreases with increasing magnetic field.} 
\label{fig:QMC_dmu}
\end{figure*}

\begin{figure*}
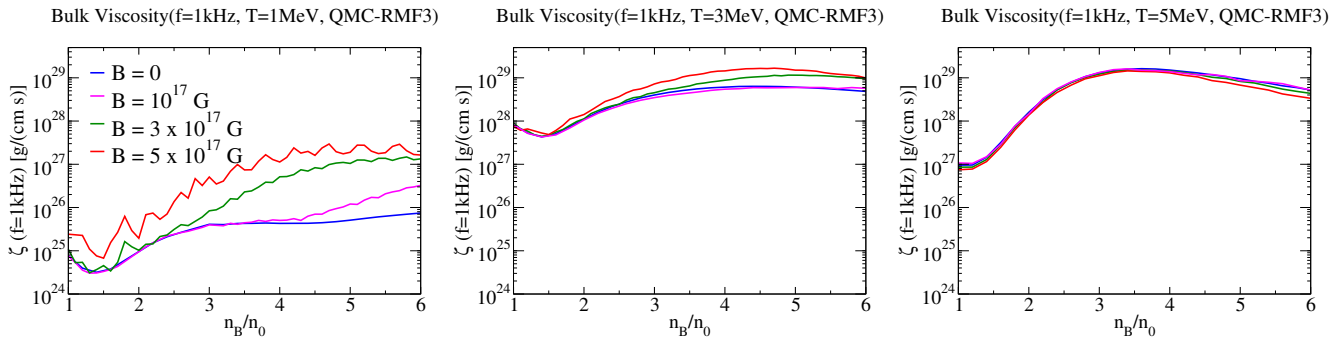

    \begin{subfigure}{0.32\textwidth}
        \centering
    \includegraphics[width=\textwidth]{bvisc_T1_QMC.eps}
    \end{subfigure}\hspace{0.01\hsize}%
    \begin{subfigure}{0.32\textwidth}
        \centering
    \includegraphics[width=\textwidth]{bvisc_T3_QMC.eps}
    \end{subfigure}\hspace{0.01\hsize}%
    \begin{subfigure}{0.32\textwidth}
        \centering
    \includegraphics[width=\textwidth]{bvisc_T5_QMC.eps}
    \end{subfigure}
    \caption{
    Bulk Viscosity for a $1\,\kHz$ density oscillation in matter described by the QMC-RMF3 EoS, at $T=1,3,5$ MeV and different magnetic fields.
    }
    \label{fig:bvisc_QMC}
\end{figure*}

\subsection{QMC-RMF3}
For comparison with our calculations of Urca rates and associated bulk viscosity for the IUF EoS, we now show results for an EoS that has no direct Urca threshold in the relevant density range, QMC-RMF3 \cite{Alford:2022bpp, Alford:2023rgp}, so it is always below the direct Urca thershold. As shown in Fig.~\ref{fig:momshort}, for QMC-RMF3 EoS the momentum deficit $k_{F_n}-k_{F_p}-k_{F_e}$ decreases with increasing density beyond $2\,n_0$, so the direct Urca process becomes less severely blocked as the density rises.

In Fig.~\ref{fig:QMC_T3_finiteB} we plot the neutron decay and electron capture rates at $T=3\,\MeV$ for $B=3\times10^{17}\,\G$ and $5\times10^{17}\,\G$.
We see that QMC-RMF3 shows a similar pattern of behavior to below-threshold IUF (Fig.~\ref{fig:IUF_ND_finiteB} and Fig.~\ref{fig:IUF_EC_finiteB}): the neutron decay rate experiences greater enhancement from the magnetic field than the electron capture rate.

As noted in Sec.~\ref{sec:results}, the mismatch between the rates leads to a temperature-driven correction $\muI$ to the flavor equilibrium condition, which we plot for QMC-RMF3 in Fig.~\ref{fig:QMC_dmu} for $T=1,3,5\,\MeV$ and different magnetic fields. We observe that at higher density, where magnetic field effects become significant for neutron decay process, $\muI$ decreases with magnetic field. This is because, as seen in Sec.~\ref{sec:results-IUF} for the IUF EoS, the field enhances
the neutron decay rate more than the electron capture rate, reducing the imbalance that $\muI$ is compensating for.

We show the plots for bulk viscosity arising from flavor equilibration by Urca processes for oscillation of frequency $f=1$ kHz at $T=1,3,5\,\MeV$ in Fig.~\ref{fig:bvisc_QMC} for different magnetic fields. Our calculations show that the isospin relaxation rate, $\gamma_I$, becomes equal to the oscillation frequency $\omega=2\pi f$ at temperatures of around $T\approx3\,\MeV$. The bulk viscosity should also peak around this temperature. For higher temperatures, the bulk viscosity should decrease with the magnetic field, as the rates increase with it.
We observe that at a low temperature $T=1\,\MeV$, the bulk viscosity increases with the magnetic field. At this temperature, the rate without a magnetic field is too slow to achieve resonance. As the Urca process rate increases with the magnetic field, the bulk viscosity thus also rises. As the temperature increases to $T=3\,\MeV$, the magnetic field effects are washed out, but we still observe a small enhancement. At $T=5$ MeV, the rate is already above resonance without the magnetic field, and we observe a small reduction, but the finite temperature effects largely overwhelm the contributions of the magnetic field.
\section{Conclusions}
\label{sec:discussions}
In this work, we have consistently calculated magnetic field effects on flavor equilibration rates and the resulting bulk viscosity in $npe$ matter at finite temperatures by evaluating the rates of equilibrating Urca processes in the NWA formalism in the presence of a magnetic field. The calculations have been performed for two finite temperature EoSs, IUF, which has a direct Urca threshold at $\ndUrca = 4.1n_0$, and QMC-RMF3 which has no direct Urca threshold.
We also studied the impact of magnetic fields on the finite-temperature correction to the flavor-equilibrium condition.

We found that, at densities below the direct Urca threshold, magnetic fields in the $10^{17}\,\G$ range noticeably enhance the rate of Urca processes. Such fields have a greater effect on neutron decay than on electron capture. This means that at those densities
the finite-temperature correction to the flavor-equilibrium condition,  $\muI = \mu_n - \mu_p - \mu_e$, decreases in the presence of a magnetic field. The magnetic field effects are swamped by thermal effects as the temperature rises above about $5\,\MeV$.

The enhancement of Urca rates means that the isospin relaxation time $\tau_I$ decreases with increasing magnetic field. We also study the impact of magnetic fields on the bulk viscous dissipation of small amplitude density oscillations of frequency $1\,\kHz$, which are relevant in merger scenarios. The bulk viscosity in matter below the direct Urca threshold density increases with magnetic field at lower temperatures when $\tau_I > 1/\omega$, since the enhancement of the isospin relaxation rate brings it closer to resonance with the $1\,\kHz$ oscillation frequency.

Consistently with this, increasing magnetic field lowers the temperature at which the bulk viscosity reaches its resonant peak, and hence the damping of density oscillations is fastest. This can lead to effective bulk viscous damping in cooler regions of a merger if the magnetic field is significant. 

Our work also serves as a demonstration of the power of the NWA formalism, which can consistently include
magnetic fields along with thermal and in-medium collisional broadening effects in flavor equilibration processes. This approach can be extended to other systems where magnetic fields and in-medium collisions play a dominant role, like color-superconducting quark matter \cite{Alford:2025jtm}, or certain beyond the standard model models, like axions, or dark baryons \cite{Harris:2025isp}.

\begin{acknowledgments}
D.C. acknowledges Liam Brodie for insightful discussions.
D.C. is thankful for the warm hospitality of Prof.
Mark Alford, Alexander Haber, and their research group
at Washington University in St. Louis. P.T. and
D.C. acknowledge the usage of IUCAA HPC computing
facility. MGA and AH are partly supported by the U.S.
Department of Energy, Office of Science, Office of Nuclear Physics, under Award No. \#DE-FG02-05ER41375. A.H.~furthermore acknowledges financial support by the UKRI under the Horizon Europe Guarantee project EP/Z000939/1.

\end{acknowledgments}

\appendix
\section{Including Landau Quantization effects in the EoS}
\label{app:landauquant}
\begin{figure}
    \centering
    \includegraphics[width=0.45\textwidth]{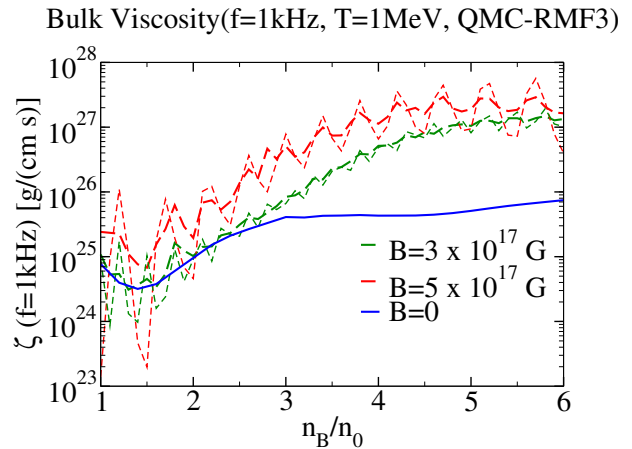}
    \caption{Bulk Viscosity for QMC-RMF3 EoS at $T=1\,\MeV$ for different values of the magnetic field. The thin dashed lines include magnetic field effects in the EoS.}
    \label{fig:bvisc_T1_QMC_magnetised}
\end{figure}

The calculations for bulk viscosity in this paper are performed using finite temperature EoSs and ignoring the magnetic field effects on the EoS itself. We here show the plot of bulk viscosity for QMC-RMF3 EoS at $T=1\,\MeV$ with the inclusion of magnetic field effects in the EoS in Fig.~\ref{fig:bvisc_T1_QMC_magnetised}. While the order of bulk viscosity is the same, the amplitude of oscillations corresponding to the increase in Landau levels is enhanced if the magnetic field effects are included in the EoS. We observe that the NWA rates and the isospin-equilibrium condition remain unchanged.

\section{Estimating contributions to below-threshold rates}
\label{app:estimating}

Here we estimate the below-threshold rates that arise from different sources: thermal blurring of the Fermi surfaces, collisional broadening of the nucleon dispersion relations, and the magnetic field.

\newcommand{\miss}{\text{miss}}
The common background is that the direct Urca process is blocked by a momentum deficit $k_\miss$, which translates to an energy deficit $E_\miss=k_\miss/v_{Fp}= k_\miss M_p/k_{Fp}$, assuming the proton is the relevant particle.
The momentum deficit for both the EoSs that we consider is plotted in Fig.~\ref{fig:momshort}.

\bigskip\noindent
(1) \uline{Thermal blurring of the Fermi surfaces} gives a contribution to the direct Urca rate that arises from the occupation of states that are thermally excited enough to overcome the missing momentum,
\begin{equation}
 \Gamma_T \approx \exp\Bigl(-\dfrac{E_\miss}{T}\Bigr)\, \Gamma_\dUrca \ 
 = \exp\Bigl( -\dfrac{k_\miss M_p}{k_{Fp} T} \Bigr)\, \Gamma_\dUrca
 \label{eq:R-est-thermal}
\end{equation}
where $\Gamma_\dUrca$ is a typical above-threshold direct Urca rate

\bigskip\noindent
(2) \uline{Collisional broadening}. This is a separate temperature-driven enhancement of the rate, which we calculate in the NWA formalism. The NWA contribution can be estimated as the integral over the tail of the Breit-Wigner function
\begin{equation}
    \Gamma_\text{NWA} \approx
    \Bigl(\int_{-\infty}^{M_p-M_\miss} R(m,M_p,W_p) \Bigr) \Gamma_\dUrca
\end{equation}
where $M_\miss$ is the smallest shift in the proton mass that will make direct Urca allowed and $W_p$ is the proton width.
Evaluating this integral,
\begin{equation}
\Gamma_\text{NWA}\approx
    \biggl( \dfrac{\pi}{2} - \tan^{-1} \Bigl(\dfrac{2M_\miss}{W_p}\Bigr) \biggr) \,\Gamma_\dUrca \ .
\end{equation}
This has the expected behavior, tending to zero when the width is too small to compensate for the momentum deficit ($M_\miss \gg W$) and tending to a factor of order 1 when the width is large enough ($M_\miss \ll W$).
In the small width (large deficit) limit we can expand in powers of $W_p/M_\miss$, 
\begin{equation}
    \Gamma_\text{NWA} \approx \dfrac{W_p}{\pt{2}M_\miss} \,\Gamma_\dUrca \ ,
\end{equation}
where
\begin{equation}
    M_\miss \approx \dfrac{dM}{dk} k_\miss =\dfrac{k_{Fp}}{M_p} k_\miss\ .
\end{equation}
Thus, 
\begin{equation}
    \Gamma_\text{NWA} \approx \dfrac{W_p M_p}{\pt{2}k_{Fp} k_\miss} \,\Gamma_\dUrca \ .
    \label{eq:R-est-NWA-W}
\end{equation}
Since $W_p = T^2/T_W$ (Eq.~\ref{eq:width}),
\begin{equation}
    \Gamma_\text{NWA} \approx \dfrac{ M_p T^2}{\pt{2}k_{Fp} k_\miss T_W} \,\Gamma_\dUrca \ .
    \label{eq:R-est-NWA}
\end{equation}
So to see when the NWA contribution is more important than thermal broadening, we compare Eq.~\eqref{eq:R-est-NWA} with Eq.~\eqref{eq:R-est-thermal}. At the lowest temperatures NWA always dominates because $T^2$ drops off slower than $\exp(-1/T)$.

\bigskip\noindent
(3) \uline{Magnetic field}.\\
To estimate the magnetic-field-induced rate below the direct Urca threshold we just need to estimate the factor $R^{qc}_B$ in Eq.~(28) in Ref.~\cite{Tambe:2024usx},
\begin{equation}
    \Gamma_B = R_B^{qc}\, \Gamma_{dU} \ ,
    \label{eq:Gamma-B}
\end{equation}
where
\begin{equation}
\begin{split}
    R_B^{qc}=&\int_{-1}^1 d\,\text{cos}\theta_p\, d\,\text{cos}\theta_e \,k_{F_p}\,k_{F_e}\frac{F^2_{l',l}(u)}{2eB} \\&\Theta (k_{F_n}-\lvert k_{F_p}\,\text{cos}\theta_p +k_{F_e}\,\text{cos}\theta_e\rvert ) \, .
     \label{eq:Rbqc_lowb}
\end{split}
\end{equation}
We can define the $x$ and $y$ parameters as in \cite{Baiko:1998jq}:
\begin{equation}
    x= \frac{k_{F_n}^2-(k_{F_p} + k_{F_e})^2}{k_{F_p}^2}N_{F_p}^{2/3}, \quad y=N_{F_p}^{2/3},
    \label{eq:def_x_y}
\end{equation}
where $N_{F_p}=k_{F_p}^2/2eB$. 
For low magnetic field 
we can use the low field asymptotic form for the Laguerre functions,
\begin{equation}
    \mathcal{F}=\frac{F^2_{l',l}}{2eB}= \frac{(\text{sin}\theta_p \text{sin}\theta_e)^{-1/3}}{y(\text{sin}\theta_p +\text{sin}\theta_e)^{2/3}}\, \text{Ai}^2(\xi)\, ,
    \label{eq:F_lowb}
\end{equation}
\begin{equation}
    \xi=[x+2y(1-\cos(\theta_p - \theta_e))]\frac{(\text{sin}\theta_p \text{sin}\theta_e)^{1/3}}{(\text{sin}\theta_p +\text{sin}\theta_e)^{4/3}}  .
    \label{eq:xi}
\end{equation}
This gives $R^{qc}_B$ below the direct Urca threshold ($k_{Fn}>k_{Fe}+k_{Fp}$) as shown in Eq.~(20) in \cite{Baiko:1998jq},
\begin{equation}
    R^{qc}_B = \sqrt{\frac{y}{x+12y}}\frac{3}{x^{3/2}}\, \exp\left(-\frac{x^{3/2}}{3}\right).
    \label{eq:R-est-mag}
\end{equation}
This gives the expected behavior that at any given below-threshold density, $R_B^{qc}$ goes to zero as the magnetic field goes to zero, since that corresponds to $y\gg 1$ and hence (via Eq.~\eqref{eq:def_x_y}), $x\gg 1$. As the density approaches the direct Urca threshold from below, we expect that even small magnetic fields will enhance the rate, and this is indeed what happens, since the constant of proportionality between $x$ and $y$ tends to zero, so $x$ can be close to 1 even when $y$ is large.

To estimate when magnetic field effects should have a  significant impact on the below-threshold Urca rate, we can compare Eq.~\eqref{eq:R-est-mag}
with \eqref{eq:R-est-thermal} to see when they outweigh thermal corrections to direct Urca, and with \eqref{eq:R-est-NWA} to see when they outweigh collisional broadening (``modified Urca'').

\clearpage  
\bibliography{bibliography}

\end{document}